\documentclass[%
 reprint,
 amsmath,amssymb,
 aps,
pre,
 longbibliography,
 lengthcheck,%
]{revtex4}

\usepackage{graphicx}
\usepackage{dcolumn}
\usepackage{bm}
\usepackage{hyperref}

\begin{document}


\title{Non-affine heterogeneities and droplet fluctuations in an equilibrium crystalline solid}
\author{Tamoghna Das$^{1}$, Surajit Sengupta$^{2,1,5}$, Madan Rao$^{3,4,5}$}
\affiliation{$^1$Advanced Materials Research Unit, S. N. Bose National Centre for Basic Sciences, Salt Lake, Kolkata 700091, India\\$^{2}$Centre for Advanced Materials, Indian Association for the Cultivation of Science, Jadavpur, Kolkata 700032, India\\$^{3}$ Raman Research Institute, C.V. Raman Avenue, Bangalore 560080, India \\$^{4}$ National Centre for Biological Sciences (TIFR), Bellary Road, Bangalore 560065, India\\$^{5}$ Kavli Institute for Theoretical Physics, UCSB, Santa Barbara, CA 93106-4030}

\date{\today}

\begin{abstract}
We show, using molecular dynamics simulations, that a two-dimensional Lennard-Jones solid exhibits droplet fluctuations characterized by {\em non-affine} deviations from local crystallinity. The fraction of particles in these droplets increases as the mean density of the solid decreases and approaches $\approx 20$\% of the total number in the vicinity of the fluid-solid phase boundary. We monitor the geometry, local equation of state, density correlations and van Hove functions of these droplets. We provide evidence that these non-affine heterogeneities should be interpreted as being droplet fluctuations from nearby, metastable minima. The local excess pressure of the droplets plotted against the local number density shows a van der Waal's loop with distinct branches corresponding to fluid-like compact,  and string-like glassy droplets.  The distinction between fluid-like and glassy droplets disappears above a well defined temperature.
\end{abstract}

\pacs{Valid PACS appear here}
\maketitle

\section{Introduction}

Crystalline solids typically exhibit local non-affine deformations when driven by external stresses. In many instances these non-affine deformations can be described in terms of a density of dislocations, however such a description is problematic when the density of dislocations is large enough that their cores overlap \cite{LL}. Since the overlapping cores of the dislocations have the character of a fluid, it has been suggested that these excitations should be thought of as fluid-like droplets \cite{Egami,Grest,Argon,spaepen}. This has proved a useful interpretation, especially since amorphous solids, for which dislocations
are difficult to define, also show such localized deformations under shear.

Just as dislocations in a solid can be thermally excited in the absence of external drive, it is reasonable to ask whether these fluid-like droplet fluctuations\cite{yukalov} can arise in the absence of external perturbation, especially when close to the fluid-solid phase boundary. Droplet fluctuations have been studied in great detail for simple Ising systems undergoing a first order transition where they are known to influence the asymptotic behavior of dynamic correlations and introduce subtle essential singularities in the equilibrium free energy\cite{Ising}.  The nature and role of droplet fluctuations in solids, on the other hand, has not received similar attention. In this paper, using a molecular dynamics (MD) simulation of a two-dimensional (2D) Lennard-Jones (LJ) solid, we show that, indeed, thermally excited droplet fluctuations do exist close to melting. We characterize the local droplet fluctuations using a non-affine order parameter  \cite{FL}, and further classify them as being fluid-like  or  ``glassy'' (reflecting the whole family of non-crystalline metastable configurations). 

\begin{figure}[h]
\begin{center}
\includegraphics[width=6.0cm]{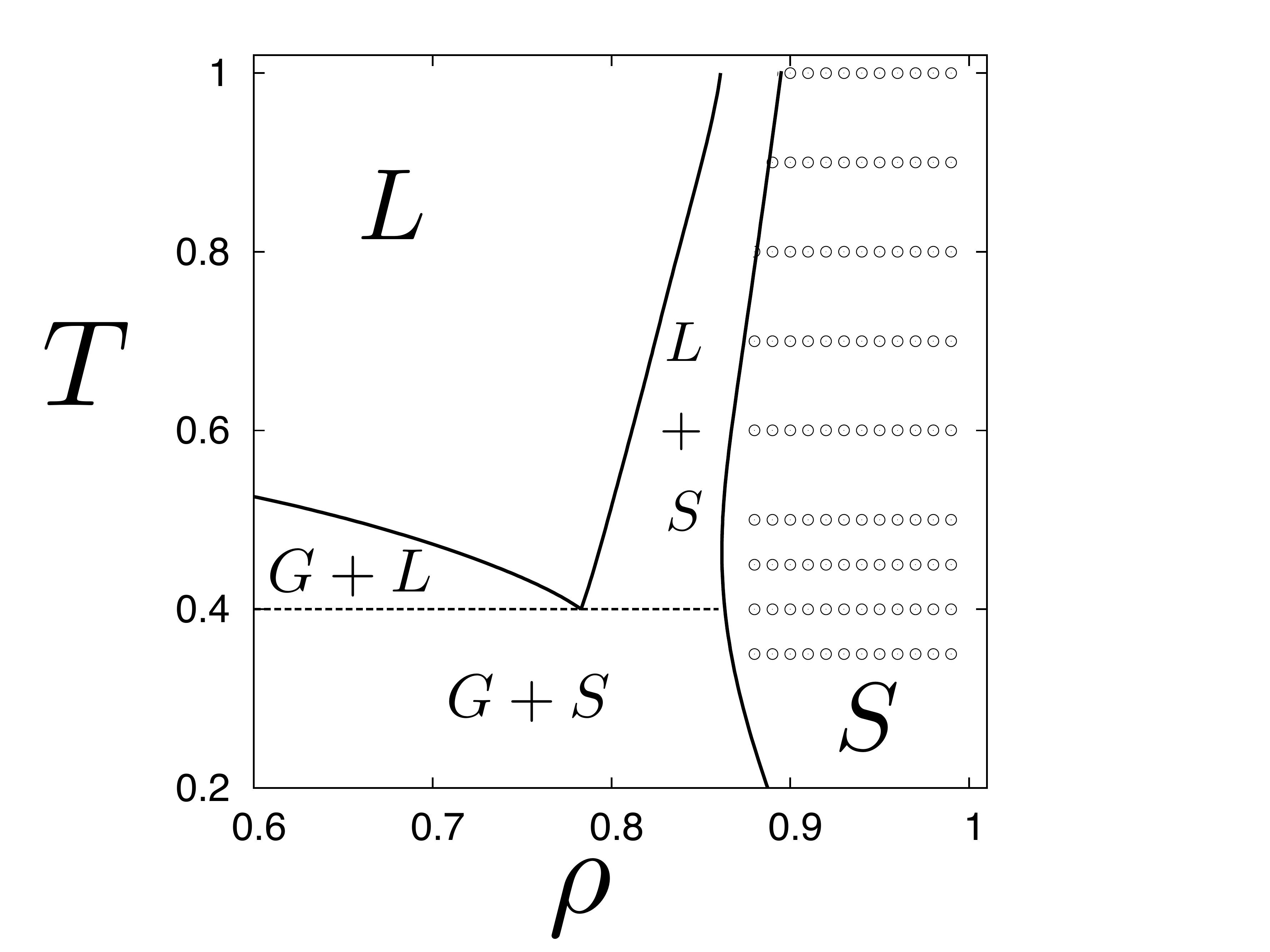}
\caption{Phase diagram of 2D Lennard-Jones solid as given in \cite{LJ}. The first order boundaries are shown by solid lines. Open circles indicate the $T$ and $\rho$ values at which we have performed MD simulations.}
\label{ljpd}
\end{center}
\end{figure}

We report MD simulations of  a 2D single-component system with the atoms interacting via LJ potential, viz.
\begin{equation}
\phi(r) = 4\epsilon\left[(\sigma/r)^{12} - (\sigma/r)^6\right]
\label{ljpot}
\end{equation}
where $\epsilon$ and $\sigma$ set the scale of energy and length whereas $\tau_0 = (m\sigma^2/\epsilon)^{1/2}$ sets the scale for time with $m$ as the mass of the particles. We use $\epsilon=\sigma=\tau_0=1$ without loss of generality. The phase diagram of this system as obtained from an earlier Monte Carlo study\cite{LJ} is shown in Fig.\ref{ljpd} in the scaled temperature $T$ - number density $\rho$ plane. First order liquid-solid and gas-solid boundaries are shown. In Fig.\ref{ljpd} we have also shown the $\rho$ and $T$ values at which we have obtained our results from equilibrated configurations.  All our state points lie in the single phase region where one always obtains an equilibrium, triangular solid. Data showing any evidence of local melting of the solid is discarded.

The main results of this paper are summarized below:
\begin{enumerate}
\item We show that there is a significant fraction of non-affine droplets in a 2D solid as the density is reduced, the fraction of particles in droplets reaches to about $20$\% at melting. 
\item The droplets are characterized by a density and excess pressure over the solid, with positive excess pressures associated with string-like droplets and negative, with compact droplets. 
\item The density fluctuations of the droplets obeys a distinct fluctuation-response relation associated with the susceptibility of the droplet. 
\item The excess pressure of a droplet of a given size, as we move across the phase diagram, depends on the shape of the droplets and is a non-monotonic function of the density at low temperatures. At high temperatures, this increases monotonically with density.
\item Finally, we show that the equal and unequal time density correlations within the droplets are liquid-like for the compact and glassy for the string-like droplets.
\end{enumerate}
Taken together, these results suggest that the non-affine droplets should be viewed as fluctuations
arising from nearby metastable liquid and glassy minima.

The rest of the paper is organized as follows. In the next section we give details of our MD simulations and the data analysis scheme we use to identify non-affine droplets. We next describe our results for the droplet shape, local thermodynamics and density correlations. Finally, we discuss the significance of our results and conclude.  

\section{Simulation and data analysis}
\begin{figure}[t]
\begin{center}
\includegraphics[width=8.0cm]{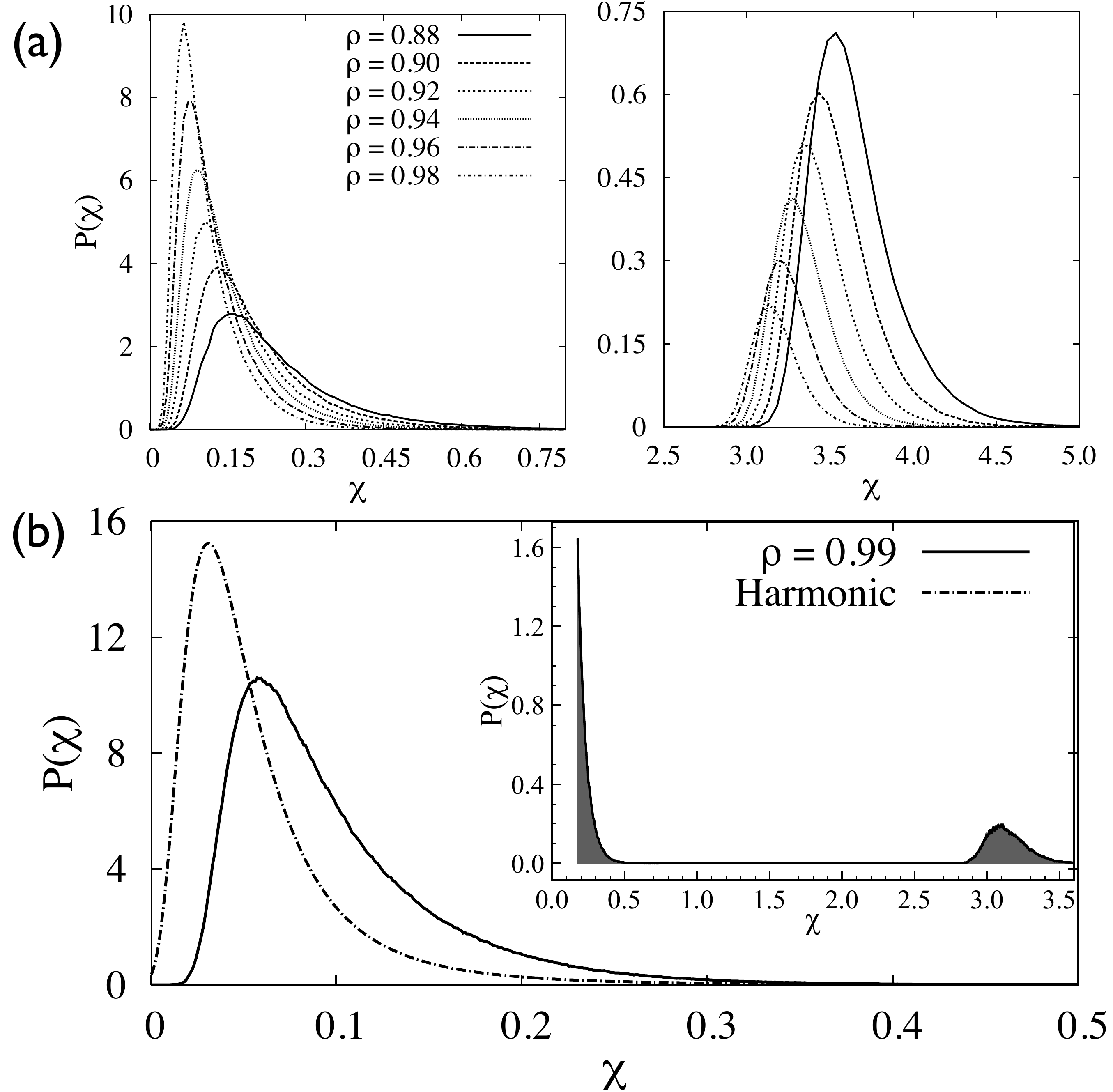}
 \caption{(a) Probability distribution of the non-affine parameter $\chi$ for $T=0.4$ at several densities. For these values of $\rho$, the $P(\chi)$ are bimodal, the contribution to the first peak is shown on the left and for the second peak on the right (note change of scale) (b) Comparison of $P(\chi)$ with that of a harmonic solid $P_{harm}(\chi)$. (Inset) Plot of $P_{drop} = P(\chi)$ for $\chi > \chi_0$ and $0$ otherwise.}
\label{chi}
\end{center}
\end{figure}

Our MD simulations are carried out both in the canonical NVT and micro-canonical NVE ensembles using a  velocity Verlet algorithm\cite{ums} with a time-step of $10^{-4}$. Starting from a system of $10^4$ particles arranged in a regular triangular lattice at desired $\rho = N/V$, we have chosen the initial velocity of each particle from a Maxwell-Boltzmann distribution at temperature, $T$. We equilibrate the system at $T$ for an initial $2\times10^5$  MD time steps. We then switch to a constant $NVE$ ensemble and collect data for another $10^5$ MD time steps, storing configurations at regular intervals. For our system, fluctuations of $T$ are of order $1$ in $10^{-4}$. 
At a fixed $T$ and $\rho$ we analyze configurations of particles using a  local measure for non-affineness ($\chi$) defined as the residual deformation of a region $\Omega$ surrounding a particle that is left over after fitting the best affine strain measured with respect to the ideal triangular lattice at $T,\rho$\cite{FL}. The neighborhood $\Omega$, defined using a cutoff distance $\Lambda$ and consisting of $n$ particles, centered around any tagged particle $0$ at ${\bf r}$ in the initial configuration is compared with that of the same particle at time $t$. We obtain the local strain $\epsilon_{ij}$ which maps as nearly as possible all the $n$ particles from the initial to the instantaneous configuration at $t$. This is done by minimizing the (positive) scalar quantity,
\begin{eqnarray}
\chi_{\Omega}({\bf r},t) & = & \sum_{n \in \Omega} \sum_i \lbrace r_n^i(t) - r_0^i(t) - \sum_j (\delta_{ij} + \epsilon_{ij}) \nonumber \\
        &  &  \times (r^j_n(0) - r^j_0(0)) \rbrace^2 
\label{nonaffine}
\end{eqnarray}
with respect to $\epsilon_{ij}$.  Here the indices $i\,{\rm and}\,j = x,y$ and $r^i_n(t)$ and $r^i_n(0)$ are the $i^{th}$ component of the position vector of the $n^{th}$ particle in the initial and instantaneous configurations respectively. Any {\em residual} value of 
$\chi_{\Omega}({\bf r},t)$, which has units of $\sigma^2$, is a measure of {\em non-affineness}. We have chosen $\Lambda = 2.5$ as our coarse graining length. We compute the probability distribution $P(\chi)$ (Fig.\ref{chi} (a) and (b)) of the coarse grained $\chi$. 

Explicit calculations show that $\chi$ is large near defects such as vacancies and dislocations which result in a change in local coordination. For $\rho >1.0$, where the solid is expected to be almost harmonic, $P(\chi)$ shows a single peak for $\chi \leq 1.0$\cite{Kerst}. As density of the system is decreased this peak becomes shorter and broader and a second peak emerges for $2.5\leq\chi\leq 5.0$. This second peak becomes more prominent as the system approaches the liquid-solid phase boundary.  Inspection of the configurations of the particles with large $\chi$ values contributing to the second peak in $P(\chi)$ show that these typically represent changes in the local topology where a pair of particles from the next near neighbor shell becomes closer than their nearest neighbors thereby increasing the local density.

To identify the truly anharmonic droplet fluctuations at given $\rho$ and $T$,  we need to subtract out contributions to $\chi$ coming from purely harmonic distortions. In order to do this we note that for a harmonic solid, $P(\chi)$ is unimodal and has a scaling form $P(\chi;T,\rho) = P(\chi k/\Lambda^2 k_B T \rho)$ where $k$ is the spring constant of the harmonic solid.  The spring constant of the reference harmonic solid was chosen to match the probability distribution of the lattice parameter in the LJ solid, obtained from the curvature of the first peak in the (angle averaged) pair distribution function $g(r)$ of the solid. The probability distribution of $\chi$ of the equivalent harmonic solid $P_{harm}(\chi)$, was multiplied by a constant till the area of the curve matched the area of the first peak of $P(\chi)$. This procedure yields a threshold $\chi_0$ above which there is no non-affineness in the reference harmonic solid and therefore any non-affiness in the LJ solid above this value must necessarily be attributed to anharmonic fluctuations. This subtraction scheme results in the distribution $P_{\rm drop}(\chi)$ of purely equilibrium droplet fluctuations in a solid at constant $\rho$ and $T$ (Fig.\ref{chi}(b)(inset)). 

\section{Results}
\begin{figure}[h]
\begin{center}
\includegraphics[width=6.0cm]{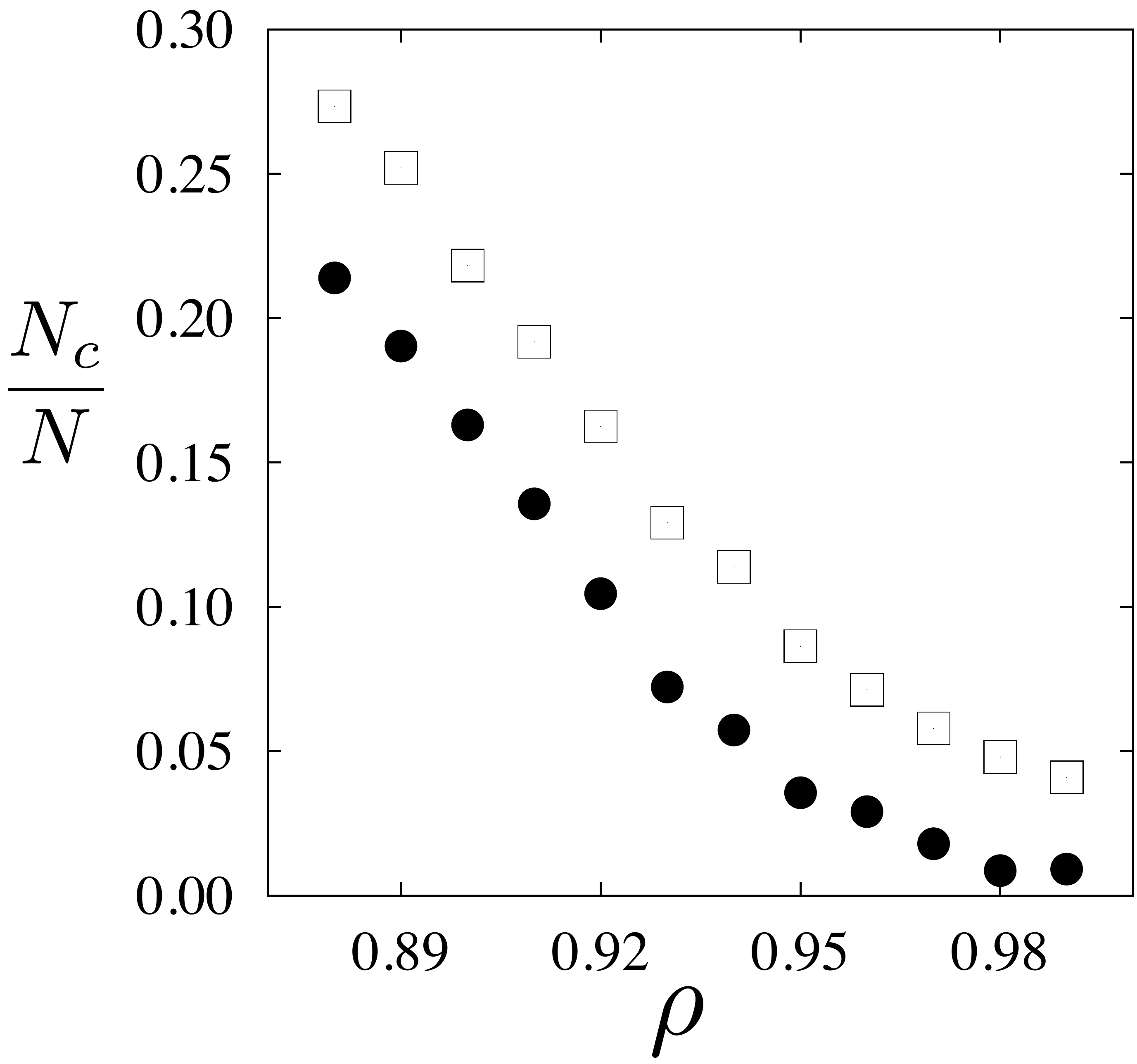}
 \caption{Plot of the fraction $N_c/N$ of particles in non-affine droplets as a function of $\rho$. The open symbols corresponds to all the non-affine particles based on the threshold criterion described in Fig. \ref{chi}(b) while the filled symbols correspond only to particles with value of $\chi$ within the second peak in $P_{\rm drop}(\chi)$. } 
\label{nc}
\end{center}
\end{figure}

We have used the above definition of non-affine particles to carry out the rest of our analysis. The fraction of particles in droplets $N_c/N$ is typically small but increases with decreasing average density, reaching approximately $20$\%  close to melting\cite{Egami} (see Fig.\ref{nc}). This might seem too large at first, however note that this is consistent with typical dislocation densities in 2d solids close to melting\cite{dis-number1,dis-number2}. We expect this droplet fraction to be much lower in 3 dimensions.  

\subsection{Droplet size and shape}
The droplets have a distribution of sizes, shapes, density and internal pressure; we compute these quantities using standard cluster counting techniques and local Delaunay analysis\cite{geom}. The number of particles in the droplets $n_c$ is exponentially distributed with a mean which increases towards the liquid-solid phase boundary. 

\begin{figure}[ht]
\begin{center}
\includegraphics[width=8cm]{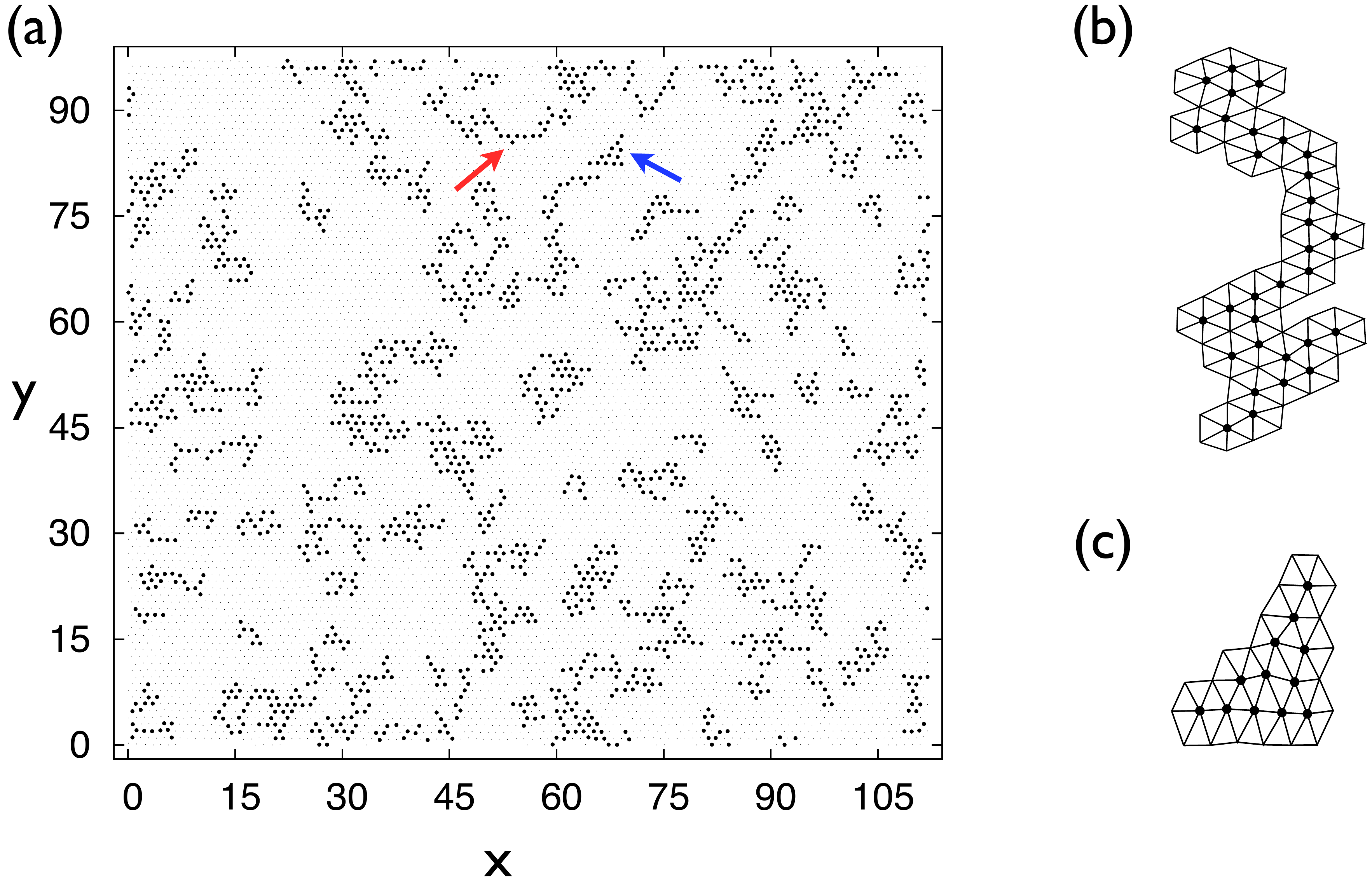}
 \caption{(a) Snapshot of a typical particle configuration in the LJ solid at $T=0.4$ and $\rho = 0.92$. The black, filled, circles denote non-affine particles while the rest of the particles are dots. A string-like and a compact droplet are pointed out using red and blue arrows respectively. Close up of a (b) string-like and (c) compact droplet showing triangulated neighborhoods. While defining droplets, we also include the nearest neighbor shell around every non-affine particle. This spatial coarse-graining improves statistics.} 
\label{snapshot}
\end{center}
\end{figure}

In Fig.\ref{snapshot}(a) we have shown a typical snapshot of the LJ particles for $\rho = 0.92$ and $T = 0.4$. To eliminate unimportant random fluctuations we show only droplets with $n_c > 7$. The snapshots show both compact and string-like morphologies.  The droplets are dynamic, they coalesce and dissociate while continuously fluctuating in shape and size. 
 
\begin{figure}[ht]
\begin{center}
\includegraphics[width=8.8cm]{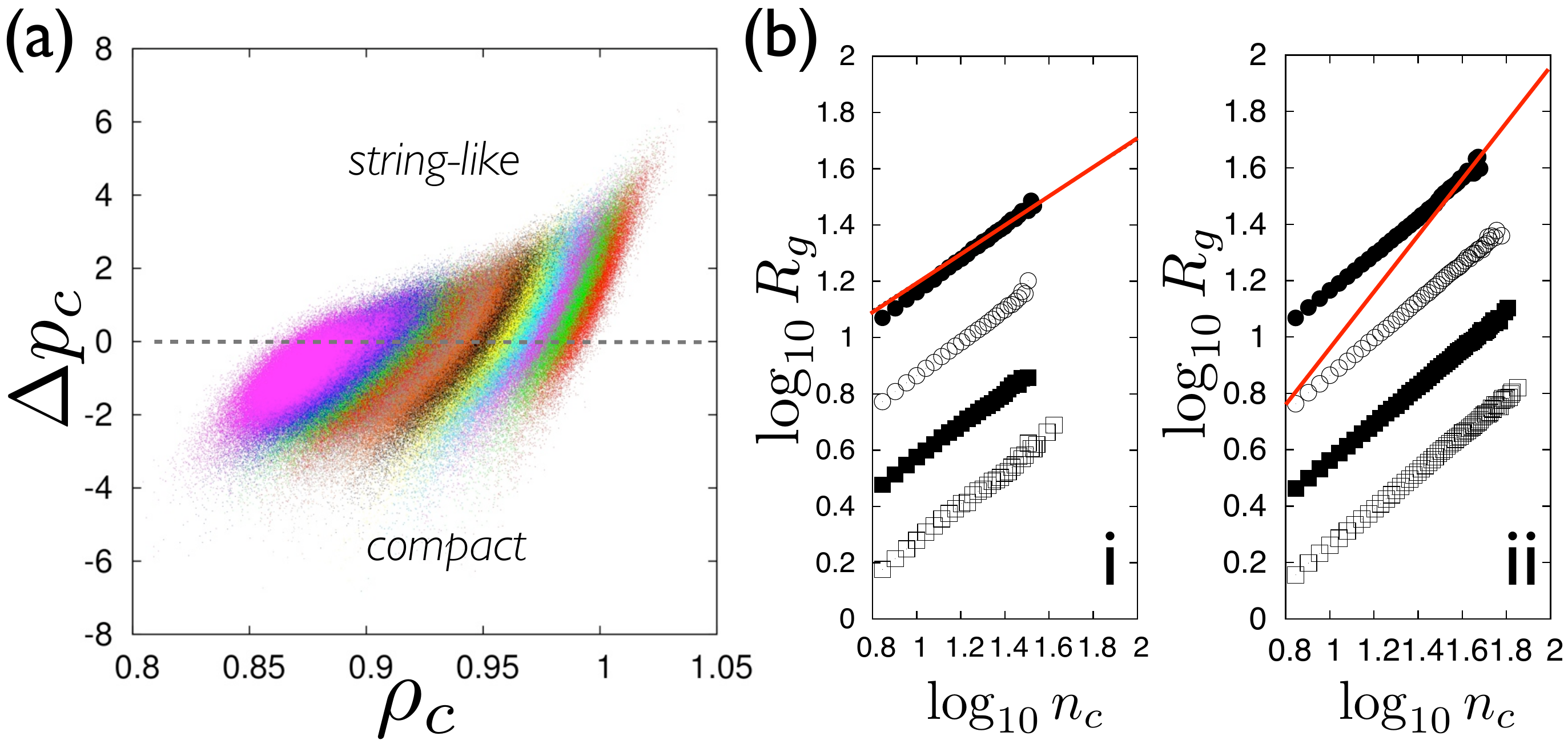}
 \caption{(color-online)(a)Excess pressure, $\Delta p_c$ vs density $\rho_c$  of the clusters as a scatter plot at densities $\rho = 0.88 - 1.0$ shown in different colors. Inspection of the individual droplets corresponding to each of the colored dots shows that for $\Delta p_c > 0$ droplets tend to be string-like while $\Delta p_c < 0$ gives rise to compact droplets. (b) The radius of gyration $R_g$ of the droplets as a function of the number of particles in the droplets $n_c$ for several $T = 0.35$ (open squares), $0.40$ (filled squares), $0.45$ (open circles) and $0.50$ (filled circles). The data points for $T > 0.35$ have been each shifted by $1$ to make them visible. Note that while $R_g \sim n_c^{1/2}$ for $\Delta p_c < 0$ (i) it is linear for droplets with $\Delta p_c > 0$ (ii) (shown, in each case by red lines).}
\label{cluster}
\end{center}
\end{figure}

To understand shape and size fluctuations of the droplets we need to obtain local densities and pressure. We do this by constructing a local Delaunay net\cite{geom} of nearest neighbor particles(see Fig.\ref{snapshot}(b) and (c)). The area of the droplet $A_c$ is then the sum of the areas of the Delaunay triangles and the density  $\rho_c\equiv n_c/A_c$. To obtain the internal pressure of the droplet, $p_c$, we compute the virial $\langle{\bf F}_{ij} \cdot {\bf r}_{ij}\rangle$ where ${\bf F}_{ij}$ and ${\bf r}_{ij}$ are the nearest neighbor forces and distances respectively for particles $i$ and $j$ belonging to the droplet. The average $\langle...\rangle$ is over all the particles $n_c$ in the droplet. The droplets are characterized by a distribution of $\rho_c$  and excess pressures, $\Delta p_c \equiv p_{c}-p$, where $p$ is the mean pressure of the surrounding solid. Figure \ref{cluster}(a) shows a scatter diagram of the excess pressure versus the density in the cluster.  We find that the droplets with high $\rho_c$ (hence large $\chi$) and  large (positive) $\Delta p_c$ are string-like, whereas droplets with low $\rho_c$ and $\Delta p_c < 0$ are compact (Fig.\ref{cluster}(a)). To study the behaviour between these two extremes, we  argue that these drops resemble 2D lattice animals with excess pressure \cite{LSF,mags}. This analogy suggests that the mean radius of gyration $R_g(n_c, \Delta p_c, T)$ obeys a crossover relation,
\begin{equation}
R_g^2 = n_c^{2\nu} F({\bar p} n_c^{2 \nu}) \,, 
\label{scaling}
\end{equation}
where ${\bar p} = \Delta p_c \sigma^2/k_B T$.
Note that our sign convention for the pressure implies that $\bar p > 0$ corresponds to {\em deflated} droplets. Further note that in \cite{LSF} the number of boundary particles are kept fixed which is a different ensembles from ours where the total number of particles in the droplets $n_c$ is fixed.

The crossover scaling function asymptotes to,
$F(x  {\rightarrow} 0)  =   {\rm const.}$ and 
$F(x \rightarrow \pm \infty)  = x^{\theta_{\pm}}$.
Note that the scaling form takes into account the natural scaling  $\bar p \sim A_c^{-1} \sim R^{-2}$ where $A_c$ is the area of the droplet. At $\bar p = 0$ we expect that the boundary of the droplet is a self avoiding random walk and hence $\nu =3/4$. The exponents $\theta_{\pm}$ take values such that $R_g \sim n_c$ for $x \to \infty$ (string-like) and $R_g \sim n_c^{1/2}$ for $x \to -\infty$ (compact).
Our data (Fig.\ref{cluster}(b)) is consistent with Eq.\ref{scaling}, however since the size of the clusters is not very large, it is difficult to probe the asymptotic behaviour. To obtain better statistics one needs to simulate yet larger systems for much longer times.

\subsection{Local thermodynamics of droplets}
The scatter diagram in Fig.\ref{cluster}(a), suggests that there might be a thermodynamic interpretation of the local density and pressure of the droplets. 
In a bulk solid, local thermodynamic equilibrium demands that the local variations in the density are related to the pressure computed from the variation $\partial F/\partial \rho$ of the Helmholtz free energy $F$ with respect to the density via the equation of state (EOS) of the solid at the ambient temperature. Further, within linear response, the generalized susceptibility $G$ obtained from the slope of the EOS is related to the $q=0$ component of the  equal-time correlation function,
$ k_B T G({\bf q}=0)  =   C({\bf q}=0,t=0) 
      =  \int d{\bf r} \langle  \delta \rho({\bf r}+{\bf x}) \delta \rho({\bf x}) \rangle
$
 where $\delta \rho$ is the deviation from the mean density\cite{BlockAna}. We check whether analogous 
 thermodynamic relations hold for the droplets taken as a subsystem in contact with the rest of the solid. We want to see how far these thermodynamic considerations apply to our configurations of droplets, with the caveat that these averages are over a restricted ensemble and therefore does not affect the equilibrium thermodynamics of the bulk solid.
 
 \begin{figure}[ht]
\begin{center}
\includegraphics[width=9.0cm]{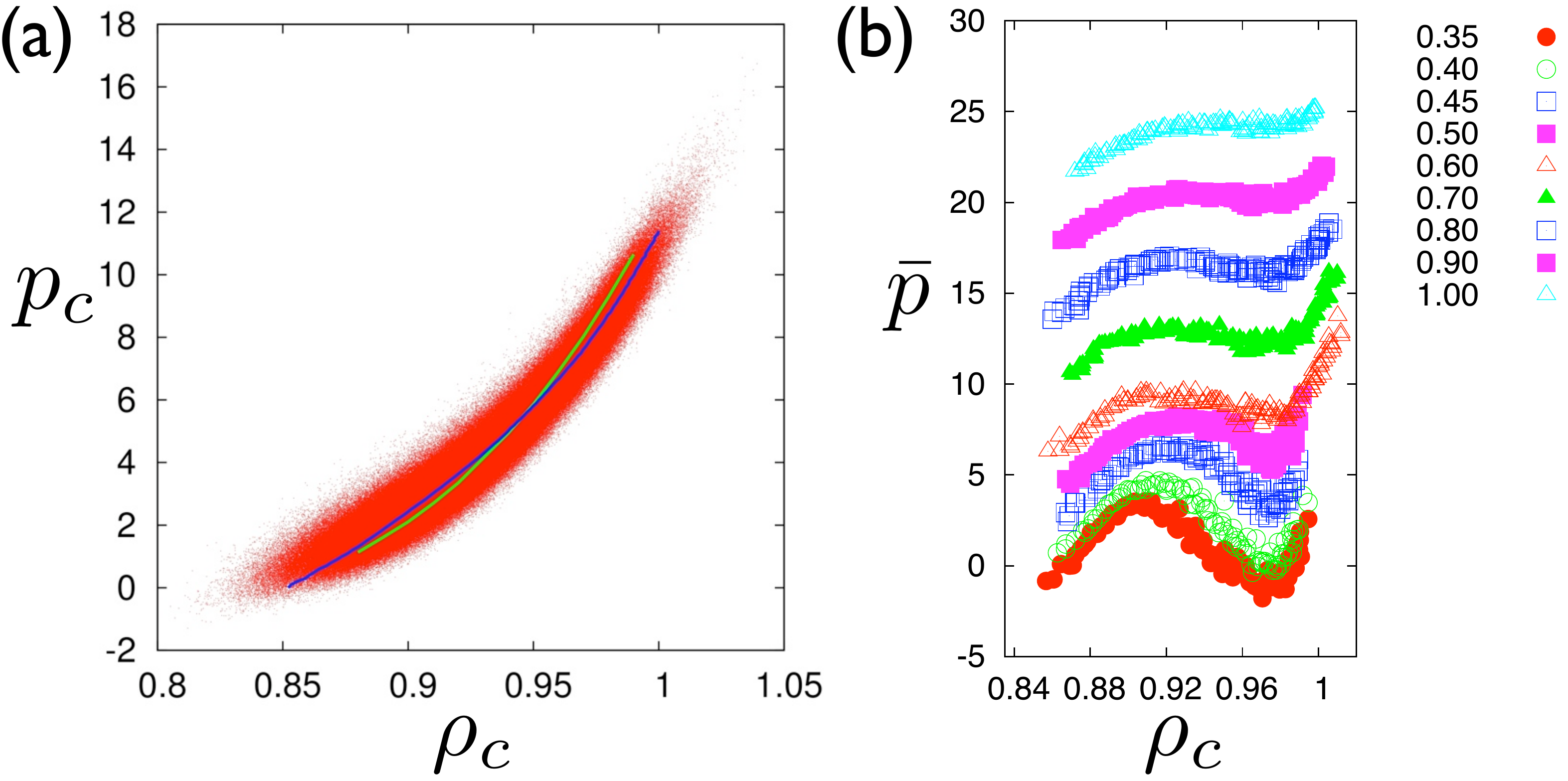}
\caption{(color-online)(a) Scatter plot of $p_c$ vs $\rho_c$ of the droplets at $T=0.40$ (red dots, the same data as in Fig.\ref{cluster}(a)) whose mean show shows a dependence similar to an ``equation of state'' (green curve). The equilibrium equation of state of the solid is shown as a blue curve for comparison.(b) The scaled excess pressure ${\bar p}$ as a function of mean density ${\bar \rho_c}$ for $n_c = 20$ particle clusters for $T=0.35-1.0$ in different colors. The data for $T > 0.35$ have each been shifted by $4$ for an unit increase in $T$ to make them visible. Note the prominent loop showing two distinct branches at low temperatures which disappears as $T$ increases. We have obtained similar data for other values $n_c$.}
\label{cleos}
\end{center}
\end{figure}
For a fixed $\rho$ and $T$ for the solid, the density $\rho_{c}$ and pressure $p_c$ of the droplets is shown as a scatter plot in  Fig.\ref{cleos}(a). The averaged curve (green line in  Fig.\ref{cleos}(a)) has a locus, ${\bar p_c}$ vs $\bar \rho_c$ which is distinct from the EOS of the equilibrium solid (blue line in Fig.\ref{cleos}(a)). 
  \begin{figure}[ht]
\begin{center}
\includegraphics[width=8.5cm]{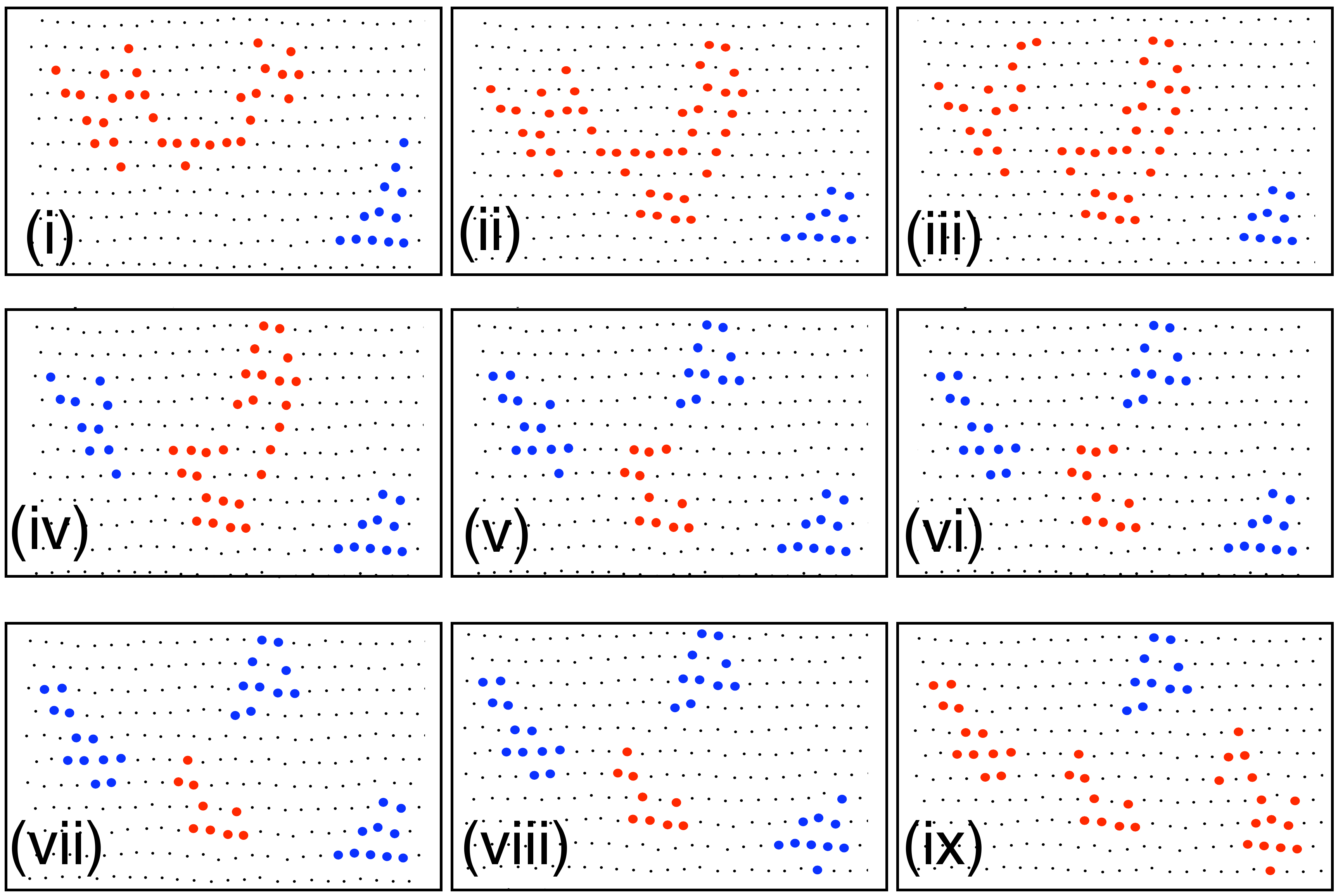}
 \caption{(color-online) Snapshots for $T=0.4$ and $\rho = 0.92$ at $0,100,150,200,250,300,350,360,$ and $370$ MD steps ((i)-(ix)) of a portion of our simulation cell showing the dynamics of two chosen droplets. The colors red and blue denotes $+$ve and $-$ve excess pressures respectively. Note that string-like droplets have $\Delta p_c > 0$ while relatively compact droplets have $\Delta p_c < 0$. Note also the dissociation of a (red) string-like droplet into a (blue) compact and a smaller (red) string-like droplet ( (iv) $\to$ (v)). } 
\label{clus-dyn}
\end{center}
\end{figure}

A plot of the scaled excess pressure ${\bar p}$ with the mean droplet density ${\bar \rho_c}$ (Fig.\ref{cleos}(b)) for fixed particle number $n_c$ shows a prominent non-linear feature akin to a van der Waals loop in the equilibrium pressure - density curve at a typical first order transition, say between gas and liquid. The two branches in the curves shown in Fig.\ref{cleos}(b) correspond to compact (liquid-like) and string-like (glassy) droplets as discussed above. Integrating this pressure - density curve  at fixed $n_c$, gives the work done by thermal fluctuations in creating a droplet of size $n_c$. We find that at low temperatures, large excess pressure tends to convert compact droplets to stringy ones and vice versa (Fig.\ref{clus-dyn}) over a characteristic relaxation time. Droplets are also seen to dissociate into distinct string-like and compact fragments with appropriate values for the excess pressure. Quite analogous to the familiar gas-liquid transition, this metastable ``van der Waals loop '' vanishes at higher temperatures beyond a  ``critical point'' which exists somewhere in the range $T = 0.9-1.0$. Above this temperature the distinction between compact and string-like droplets disappears. 
 \begin{figure}[ht]
\begin{center}
\includegraphics[width=7.0cm]{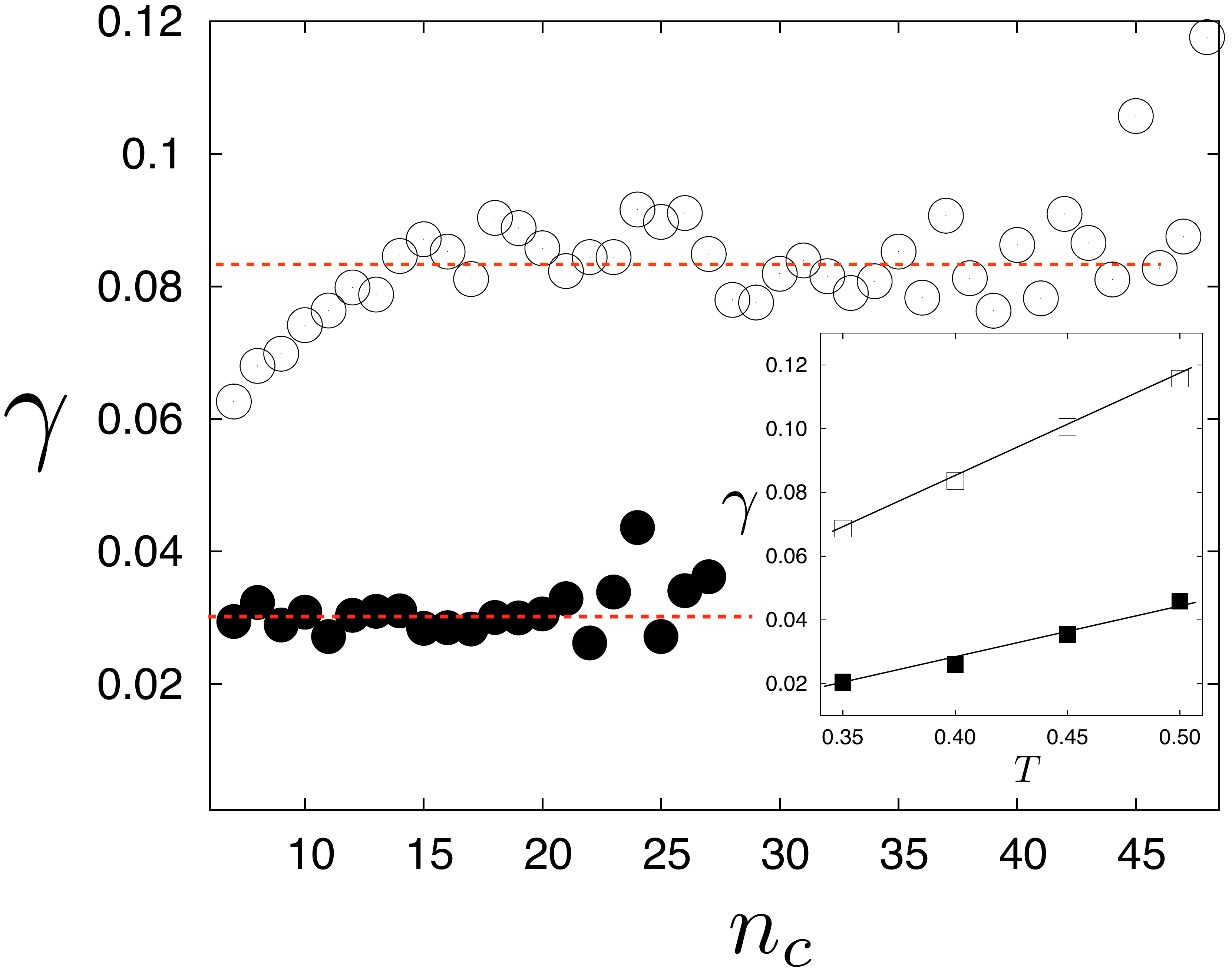}
\caption{The fluctuation -response ratio $\gamma$ shown as a function of $n_c$ for droplets with $+$ve (filled circles) and $-$ve excess pressure, $\bar p$ is independent of $n_c$, with an intercept $\propto T$. We verify this by plotting the intercept versus $T$ for all the droplets (inset) for $T\leq0.5$. The filled and unfilled squares correspond to the $+$ ve and $-$ ve $\bar p$ branches respectively.}
\label{clfr}
\end{center}
\end{figure}
We find that the generalized susceptibility $G$ obtained from the slope of $\partial \bar p_c /\partial \bar \rho_{c}$ is {\em proportional} to fluctuations of the density $\langle \delta \rho_c^2\rangle = \langle (\rho_{c} - {\bar \rho_{c}})^2\rangle$ where $\langle \cdots \rangle$ denotes a time average. The fluctuation response ratio $\gamma= \langle \delta \rho_c^2\rangle n_c/\left( \rho_c^2 \partial \bar p_c/\partial {\bar \rho_c} \right)$ should be independent of $n_c$, with an intercept which should be proportional to the temperature $T$. This is shown in Fig.\ref{clfr} where $\gamma$ for the $+$ve and $-$ve $\bar p$ branches in Fig.\ref{cleos}(b) are shown separately using filled and open circles respectively. While $\gamma$ is indeed a constant, the intercept plotted versus $T$ for all the droplets for the two branches are approximately linear in $T$. This is the metastable analogue of the equilibrium fluctuation-response relation discussed above, suggesting that the droplets are fluctuations from a metastable state describable by a free energy functional.

\subsection{Density correlations}
\begin{figure}[ht]
\begin{center}
\includegraphics[width=7.0cm]{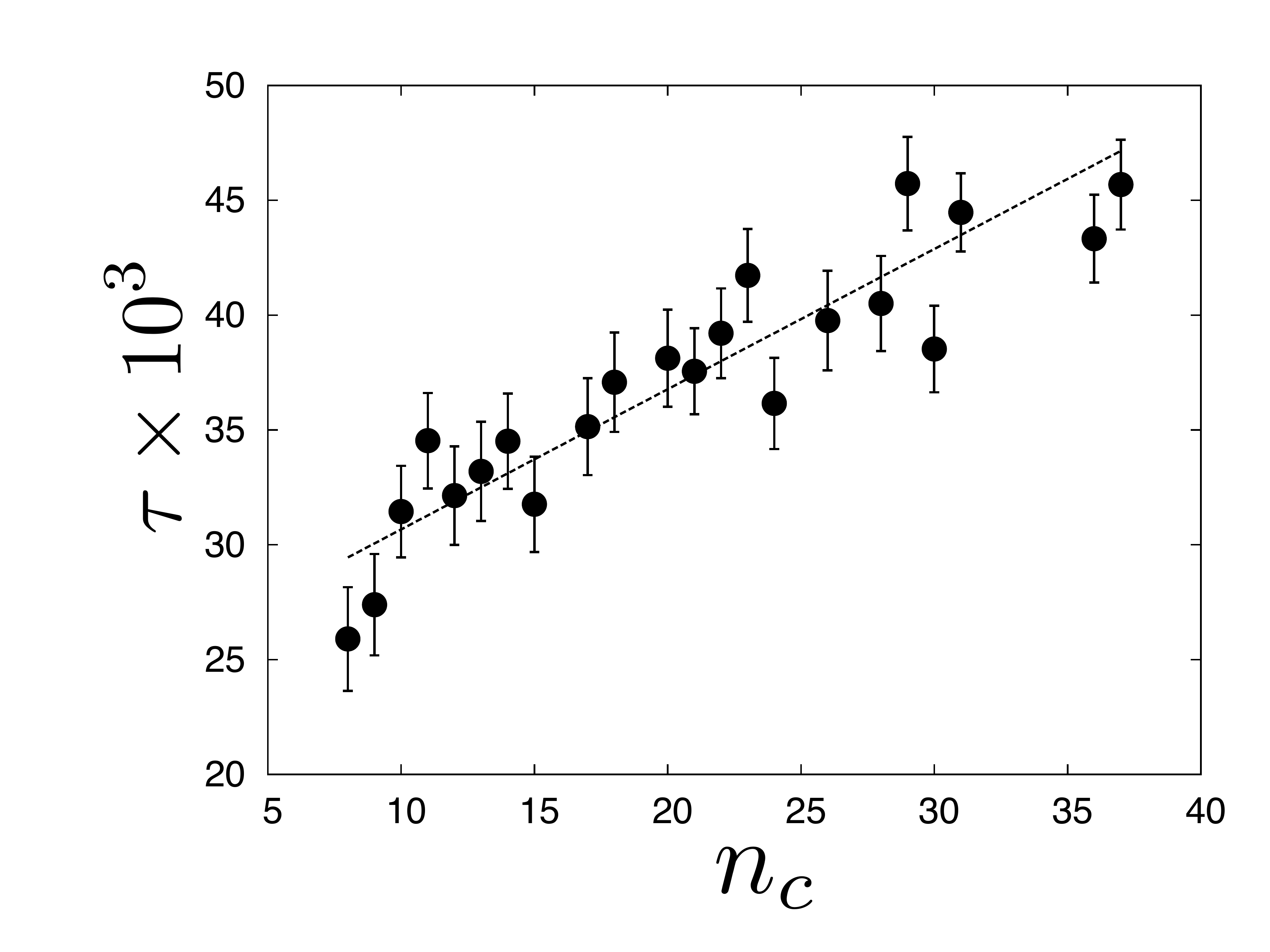}
\caption{Lifetime $\tau$ vs $n_c$ for droplets at $\rho = 0.92$ and $T=0.4$. Larger droplets survive longer. The error bars are obtained from the width of the distribution of $\tau$. The dashed straight line is a guide to the eye.}
\label{lftm}
\end{center}
\end{figure}

\begin{figure}[ht]
\begin{center}
\includegraphics[width=7.0cm]{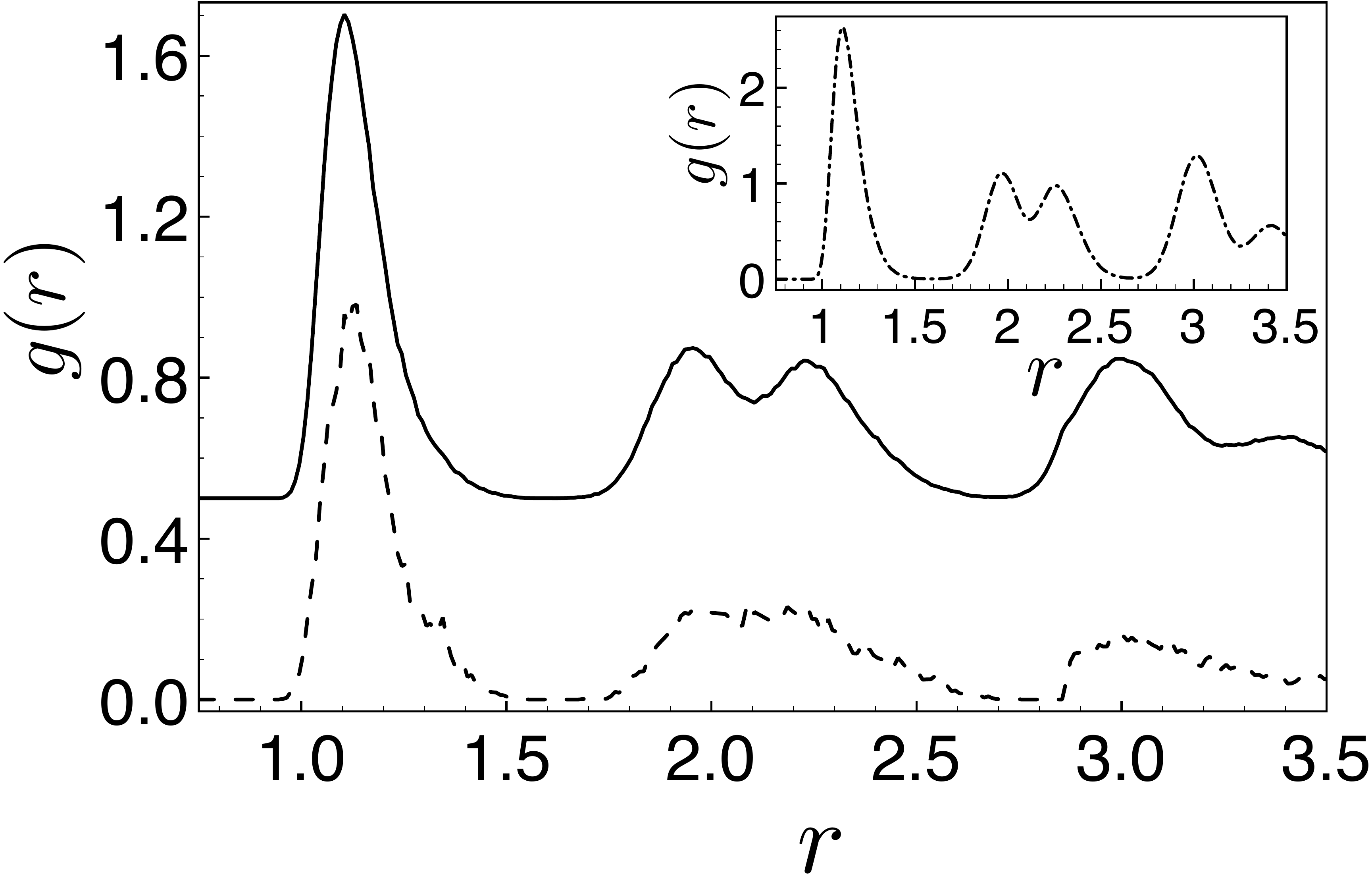}
\caption{Typical pair distribution function $g(r)$ for particles in non-affine droplets at $\rho=0.88$ and $T=0.40$, drawn separately for $\Delta p_c >0$ (solid line) and $\Delta p_c < 0$ (dashed line). The $g(r)$ for the high density droplet has been shifted by $0.5$ to make it visible. Both the $g(r)$'s show less crystalline structure than the solid (inset).}
\label{gofr}
\end{center}
\end{figure}

\begin{figure}[ht]
\begin{center}
\includegraphics[width=8.0cm]{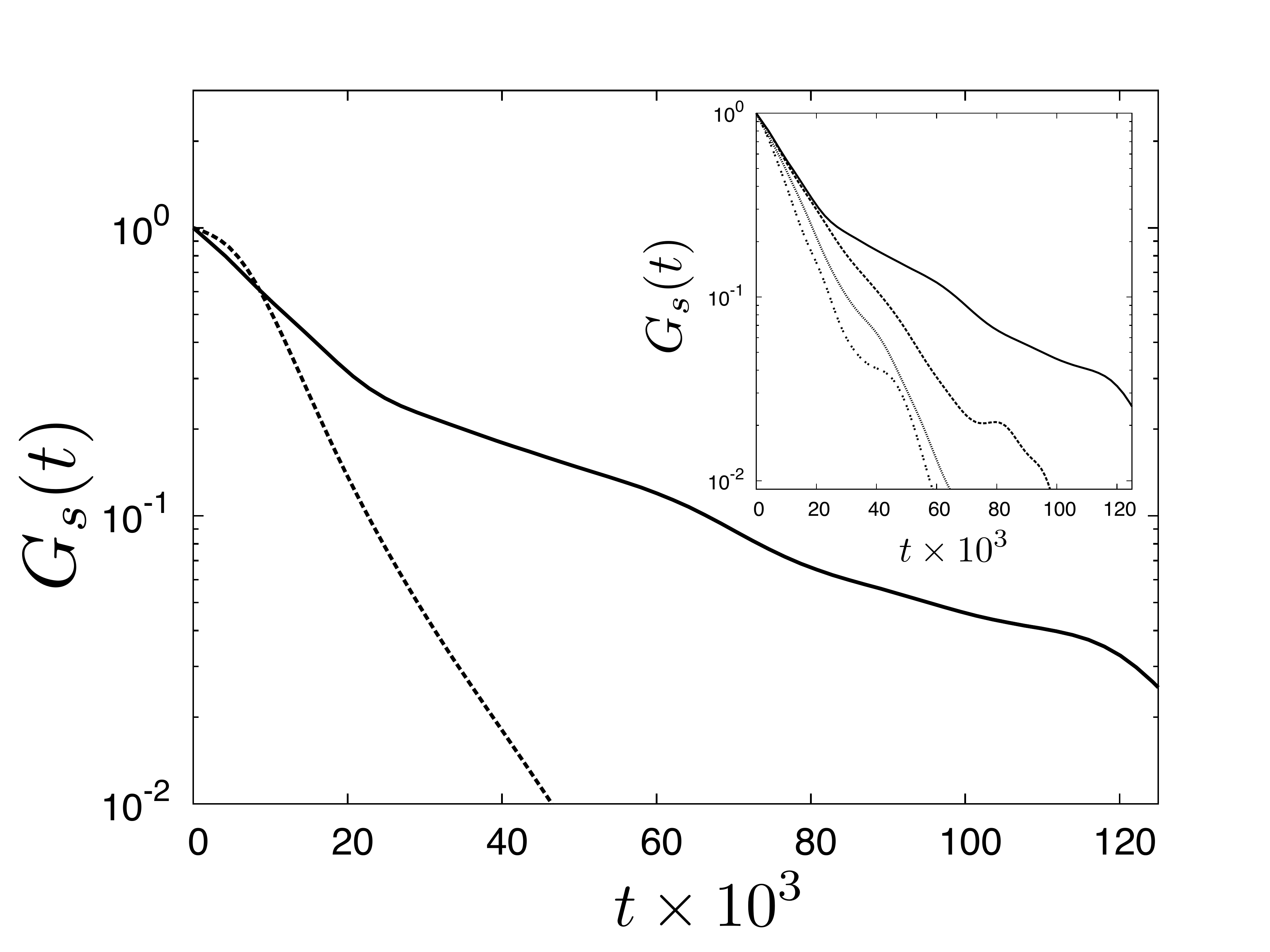}
\caption{(a) The self part of the van-Hove function $G_s(t)$ as a function of time for the same droplets whose $g(r)$ is given in Fig. \ref{gofr}. Note the prominent plateau for the high density droplets which is cut off by the finite lifetime of the droplet. (b) Development of the plateau in $G_s(t)$ with the lifetime of the droplets over which $G_s(t)$ is averaged (lines with mean $\tau$ from $50$ (bottom) to $200$ (top) $\times 10 ^{-3}$ LJ units). }
\label{vhove}
\end{center}
\end{figure}

We now study density correlations within each droplet to further characterize compact and string-like droplets. Note that these droplets have a finite lifetime $\tau(n_c)$, a plot of $\tau$ vs. $n_c$ for fixed $\rho$ and $T$ is shown in Fig.\ref{lftm}. To obtain $\tau$, we have collected the times each constituent particle continues to belong to a droplet. As the droplet fluctuates, particles from the periphery continuously attach to and detach from the droplet leaving a set of particles at the core intact. This is clear from the snapshots shown in Fig.\ref{snapshot}. We define $\tau$ as the persistence time of these core particles. Our analysis for different $\bar p$ shows that the dense stringy clusters live longer. To obtain good statistics for the equal and unequal time density correlators, we therefore need to look at large and long-lived droplets. 
Fig.\ref{gofr} shows the equal time density correlations, $g(r)$ . First of all $g(r)$ in both branches show features associated with an amorphous structure with a first peak value which is much reduced from that of the full solid. Note that while the low density droplets are more liquid-like with smoothened peaks, the high density droplets are glassy showing a prominent split second peak.
We have also computed the self part of the van Hove correlation functions $G_s(t)=\langle \rho(0,0) \rho(0,t) \rangle$. This is shown in Fig.\ref{vhove}. While $G_s(t)$ relaxes exponentially for the low density, liquid-like droplets, the high density droplets show non-Debye relaxation. We show that as the lifetime of the droplets increase, a prominent $\beta$-relaxation type plateau begins to develop. This, however, gets cut-off by the finite lifetime of the droplets in the solid. 

\section{Discussion and Conclusions}
We have shown in this paper that the excitations of an equilibrium solid at high temperatures can be interpreted as arising from a distribution of non-affine droplets whose mean size and lifetime increases as one approaches the liquid-solid phase boundary.  These droplets are characterized by a density, internal pressure and shape. The shape of the droplets crossover from being compact to string-like as the density increases.  The observed relationship between the local pressure and density of the droplets in the form of an ``equation of state'' and the fluctuation-response relation of the local density strongly suggests that these non-affine droplets arise as fluctuations from a metastable liquid or glass. Consistent with this we find that the high density droplets have a $g(r)$ and van Hove function resembling that of glasses. Our main concern in this paper has been to provide evidence that droplet fluctuations of metastable liquid or glass exist in a crystalline solid and to characterize their shape and local thermodynamic parameters. How do these fluctuations influence the properties of solids? We hope to systematically study and answer this question in the future. Some of the specific areas where the impact of droplet fluctuations may be observable are discussed below.

For example, we wonder whether our results hint at the presence of a metastable liquid-glass critical point. While Fig.\ref{cleos}(b) is certainly suggestive, we must remember that the size of the non-affine droplets are typically small with $n_c \sim 100$ even for the largest droplets. A careful finite size scaling analysis needs to be carried out in order to determine whether this feature survives for larger droplet sizes. The nature of the melting transition in the two-dimensional LJ system remains unclear in spite of being the subject of many investigations\cite{tox1,LJ,ud,chen} over several decades. Early simulations obtained a first order liquid-solid and gas-solid transitions with prominent coexistence regions\cite{tox1,LJ}.  While at low temperatures near the triple point ($T=0.4$), one obtains a first order melting transition\cite{tox1}, at higher temperatures ($T \sim 1$) one obtains a much reduced coexistence region\cite{ud} with some characteristics of continuous\cite{strand} melting. At still higher temperatures, the melting transition appears to be unequivocally driven by dislocation unbinding\cite{chen}. Exactly how and at what temperature one obtains this change in the nature of melting is, as yet, unknown. While we do not address this question in the present paper, we speculate that droplet fluctuations may have a strong influence on the dynamics of melting. The presence of metastable critical points is known to crucially influence the dynamics of first order transitions, for eg. the important problem of protein crystallization\cite{prot,sri}. 

We expect these non-affine droplets to also play an important role in the rheology of solids under applied stresses; our preliminary work in this direction is consistent with this expectation. We would also like to enquire whether it is possible to observe these droplet excitations in a real experimental situation in two and three dimensions. Such excitations, if they exist, may be difficult to disentangle from the contributions coming from a density of dislocations and grain boundaries. Perhaps direct visualization of droplet fluctuations in colloidal crystals is the best way to study these effects\cite{zahn}.

\acknowledgements
We thank A. Paul, S. Yip, P. Sollich, T. Egami, C. Chakraborty, M. Falk, F. Spaepen and C. Dasgupta for discussions and a critical reading of the manuscript. Some of the authors (SS and MR) are pleased to acknowledge that part of this research was performed while in residence at the Kavli Institute for Theoretical Physics, and was supported in part by the National Science Foundation under Grant No. PHY05-51164.


\begin{thebibliography}{}
\bibitem{LL} L. D. Landau and E. M. Lifshitz in {\it Theory of Elasticity}, (Pergamon Press,
$3^{rd}$ edition, 1986).

\bibitem{Egami} T. Egami {\em et al.}, \prb {\bf 76}, 024203 (2007).

\bibitem{Grest} M. H. Cohen and G. Grest, \prb, {\bf 20}, 1077 (1979).

\bibitem{Argon} A. Argon, Acta Met. {\bf 27}, 47 (1979).

\bibitem{spaepen} F. Spaepen, Acta Met. {\bf 25}, 407 (1977).

\bibitem{yukalov} For reviews see, K. Binder, Rep. Prog. Phys. {\bf 50}, 783, (1987); V. I. Yukalov, Phys. Rep. {\bf 208}, 395, (1991).

\bibitem{Ising} D. A. Huse and D. S. Fisher \prb, {\bf 38}, 6841, (1987); C. Tang, H. Nakanishi and J. S. Langer, \pra {\bf 40}, 995, (1989); J. S. Langer, Annals of Physics {\bf 281}, 941 ( 2000 ).

\bibitem{LJ} J. A. Barker, D. Henderson and F. F. Abraham, Physica {\bf 106A}, 226 (1981).

\bibitem{ums} D. Frenkel and B. Smit, {\em Understanding Molecular Simulations} (Academic Press, San Deigo, 2002).

\bibitem{FL}  M. L. Falk and J. S. Langer, \pre {\bf 57}, 7192 (1998).

\bibitem{Kerst}  K. Franzrahe, P. Nielaba and S. Sengupta, \pre {\bf 82}, 016112 (2010).

\bibitem{geom} We have used the GEOMPACK code downloaded from (\verb1http://people.sc.fsu.edu/~jburkardt/f_src/geompack1)

\bibitem{LSF} S. Leibler, R. R. P. Singh and M. E. Fisher, \prl {\bf 59}, 1989 (1987).

\bibitem{mags} A. C. Maggs, S. Leibler, M. E. Fisher and C. Camacho, \pra {\bf 42}, 691 (1990). 

\bibitem{dis-number1}B. Jo{\' o}s, ``The Role of Dislocations in Melting'', Chap. 55 in {\it Dislocations in Solids}, Vol. 10, F. R. N. Nabarro and M. S. Duesbery, eds. (Elsevier, Amsterdam, 1996), p.505-594. 

\bibitem{dis-number2} S. Sengupta, P. Nielaba and K. Binder, \pre {\bf 61}, 6294 (2000).

\bibitem{BlockAna} M. Rovere, P. Nielaba and K. Binder, Z. Phys. {\bf 90}, 215 (1993).

\bibitem{tox1} S. Toxvaerd, \prl, {\bf 44}, 1002 (1980).

\bibitem{ud} C. Udink and D. Frenkel, \prb, {\bf 35}, 6933 (1987).

\bibitem{chen} K. Chen, T. Kaplan, M. Mostoller, \prl , {\bf 74}, 4019 (1995).

\bibitem{strand} K. J. Strandburg, Rev. Mod. Phys. {\bf 60}, 161 (1988).

\bibitem{prot} P. R. ten Wolde and D. Frenkel, Science, {\bf 277}, 1975 (1997).

\bibitem{sri}P. Kumar, L. Xu, Z. Yan, M. G. Mazza, S. V. Buldyrev, S. H. Chen, S. Sastry, H. E. Stanley \prl, {\bf 97}, 177802

\bibitem{zahn} K. Zahn, A. Wille, G. Maret, S. Sengupta, P. Nielaba, Phys. Rev. Lett. {\bf 90}, 155506 (2003).

\end{thebibliography}
\end{document}